\documentclass[twocolumn]{emulateapj}
\usepackage{graphicx,amssymb,color}
\usepackage[fleqn]{amsmath}
\newcommand{\mytilde}{\raise.17ex\hbox{$\scriptstyle\mathtt{\sim}$}}

\begin{document}

\title{Dynamical and biological panspermia constraints within multi-planet exosystems}

\author{Dimitri Veras$^{\dag}$}\thanks{$^{\dag}$STFC Ernest Rutherford Fellow}
\affil{Centre for Exoplanets and Habitability, University of Warwick, Coventry, CV4 7AL, UK}
\affil{Department of Physics, University of Warwick, Coventry, CV4 7AL, UK}
\email{d.veras@warwick.ac.uk}

\author{David J. Armstrong}
\affil{Centre for Exoplanets and Habitability, University of Warwick, Coventry, CV4 7AL, UK}
\affil{Department of Physics, University of Warwick, Coventry, CV4 7AL, UK}

\author{James A. Blake}
\affil{Centre for Exoplanets and Habitability, University of Warwick, Coventry, CV4 7AL, UK}
\affil{Department of Physics, University of Warwick, Coventry, CV4 7AL, UK}

\author{Jose F. Guti\'{e}rrez-Marcos}
\affil{Centre for Exoplanets and Habitability, University of Warwick, Coventry, CV4 7AL, UK}
\affil{School of Life Sciences, University of Warwick, Coventry, CV4 7AL, UK}

\author{Alan P. Jackson}
\affil{Centre for Planetary Sciences, University of Toronto at Scarborough, Toronto, ON, Canada}
\affil{School of Earth and Space Exploration, Arizona State University, AZ, USA}

\author{Hendrik Sch\"{a}efer}
\affil{Centre for Exoplanets and Habitability, University of Warwick, Coventry, CV4 7AL, UK}
\affil{School of Life Sciences, University of Warwick, Coventry, CV4 7AL, UK}

\begin{abstract}
As discoveries of multiple planets in the habitable zone of their parent star mount, developing 
analytical techniques to quantify extrasolar intra-system panspermia will become increasingly important.
Here, we provide user-friendly prescriptions that describe the asteroid impact characteristics which 
would be necessary to transport life both inwards and outwards within these systems within a single framework. 
Our focus is on projectile generation and delivery and our expressions are algebraic,  eliminating the need for the 
solution of differential equations.
We derive a probability distribution function for life-bearing debris to reach a planetary orbit, and describe the survival of
micro-organisms during planetary ejection, their journey through interplanetary space, and atmospheric entry.
\end{abstract}


\section{Introduction}

Although studies of the transport of life-bearing rocks between planets have a long history \citep[e.g.][]{melosh1988}, 
claims of the discovery of traces of ancient life in the meteorite ALH84001 in the mid-1990s \citep{mcketal1996} accelerated 
investigations into panspermia within the Solar system. The last two decades have since featured detailed work \citep[e.g.][]{miletal2000} 
outlining delivery dynamics \citep{melton1993,glaetal1996,glaetal1997,alvetal2002,reyetal2012,woretal2013}, impact physics and chemistry 
\citep{meyetal2011,rubin2015,baretal2016}, and 
biological survival requirements \citep{horetal1994,masetal2001,nicholson2009,moeetal2010} with
respect to Earth, Mars and other solar system bodies. Consequently, a detailed foundation for 
panspermia-related processes has been established.

Despite these advances, the applicability of these processes to extrasolar planetary systems is still in question,
partly because in those systems we lack the detailed knowledge of our own planetary system.
Nevertheless, efforts to characterize panspermia between different extrasolar systems, or between
the solar system and extrasolar systems, have contributed to our understanding
\citep{adaspe2005,valetal2009,wesson2010,beletal2012,linloe2015,galwan2017}.
However, panspermia amongst extrasolar planets within the same system
has received a relatively little but increasing amount of attention
\citep{helarm2014,steli2016,krietal2017,linloe2017a}. A potential reason 
for this relative dearth of studies is the lack of observational evidence of 
multiple planets in the habitable zone of the same star.

This situation has now changed with the groundbreaking discovery of multiple potentially
habitable planets in the TRAPPIST-1 system \citep{giletal2017}. TRAPPIST-1 is an M dwarf
with a mass of $0.08M_{\odot}$ that harbours seven observed transiting planets all with masses 
similar to the Earth and three (planets e, f and g) which are securely in the star's habitable
zone (although all seven may be, with effective temperatures ranging from about 150K to 400K). 
Because all seven planets are seen transiting from the Earth, their orbits are nearly coplanar.
The system is compact (all planets could fit well within Mercury's orbit),
and are likely to be resonantly interacting in a long chain \citep{lugetal2017}.
We do not yet know if an equivalent of the Late Heavy Bombardment event (believed to have
occurred in the Solar system, even if less intense than originally thought; \citealt*{botnor2017}) 
has or will occur in that system,
nor what types of potential impactors lurk beyond the most distant planet (h) and outside of our
field of view. Despite the uncertainties, investigation of panspermia from within this system
has already been undertaken \citep{krietal2017,linloe2017a} and might be prompted further by additional
observations, which are currently underway.

Here, we study and derive several aspects of lithopanspermia in more general closely-packed multi-planet 
systems, with a focus on analytics and dynamical delivery, but also addressing micro-organism survival 
at each stage.
Numerous Solar system studies have 
taught us that $N$-body simulations are both computationally expensive and dependent on a large
number of parameters \citep{donetal1999} which are unknown in exoplanetary systems. Computational times 
for most known extrasolar planetary systems would be worse because of their compact nature. Therefore,
we adopt a purely analytical approach, one which could be applied to extrasolar systems
with multiple habitable planets. The characterisation of such systems is expected to increase
steadily over the next decade, culminating with the PLATO mission \citep{rauetal2014}, which will 
measure habitable planets out to about 1 au.

Throughout the paper, our subscript convention for physical quantities will be: 
$i$ for the impactor, no subscript for the source planet, a single prime for 
the fragmented debris ejected from the source planet, and a double prime for the
target planet.
In Section 2, we establish our setup and describe how a life-bearing rock could be
transferred between one planet and the orbit of another planet; Appendix A provides most
of the intermediate equations required for this section, and Appendix B contains an extension
with a fictitious template compact system which can be used for quick estimates. Section 3 then details the likelihood
of that rock actually impacting a target planet. Section 4 constrains the characteristics of the
ejecta that would both satisfy the dynamics and have the capability to harbour life. In Section 5,
we consider the biological prospects of life surviving all aspects of lithopanspermia. We conclude in Section 6.

\section{Orbit transfers}

\subsection{Setup}

Consider a pair of planets such that one, the ``source'', contains a life-bearing organism, whereas the other, the ``target'', initially does not.  An impactor crashes into the source, producing a spray of life-bearing ejecta. By assuming that the ejecta is ``kicked'' impulsively, we estimate its direction and speed such that it would reach the orbit of the target. By impulsively, we refer to the timescale of the kick being much smaller than the orbital period of the source. In multi-planet systems within the detectability threshold of transit photometry surveys, planet orbital periods are on the order of days, whereas impact kick timescales would typically be on the order of minutes.

The underlying formalism we use was established in \cite{jacwya2012} and expanded upon in \cite{jacetal2014}, and similar to that in Appendix A of \cite{feuwie2008}. We briefly repeat here the geometrical setup in \cite{jacetal2014}: Denote the kick speed as $\Delta v$, and the circular speed of the source as $v_{\rm k} \equiv \sqrt{G\left(M_{\rm star} + M\right)/a}$, where $a$ and $M$ are the source's semimajor axis and mass. The launch speed $v_{\rm lau}$ and escape speed $v_{\rm esc}$ are related to $\Delta v$ through 
$v_{\rm lau}^2 = \left(\Delta v\right)^2 + v_{\rm esc}^2$, where $v_{\rm esc}^2 = 2GM/R$.

For perspective, the circular speed of the TRAPPIST-1 planets are, from planet b to h moving outward, $\left\lbrace 79.9, 68.2, 57.5, 50.2, 43.7, 39.6, 33.9\right\rbrace$ km/s. The ratio $\Delta v / v_{\rm k}$ features frequently in the equations, and is bound from above in our study by the value of $\sqrt{2} + 1$, which is the maximum possible value for which the debris can remain in the planetary system. Further, the minimum value of $\Delta v / v_{\rm k}$ for which the ejecta can escape the system is $\sqrt{2} - 1$.

At impact, the source is assumed to lie at the pericentre of its orbit such that its argument of pericentre, longitude of ascending node, and inclination $I$ are all zero\footnote{The restrictiveness of the assumption of impact at orbital pericentre will be removed later when we assume circular orbits.}. The source is assumed to move counterclockwise, and the kick direction is defined by two variables: $\theta$ and $\phi$. The angle between the source's angular momentum vector and the kick direction is $\theta$ and the angle between the star-source pericentre line and the projection of the kick direction onto the source's orbital plane is $\phi$.

\begin{figure*}
 \includegraphics[width=9cm]{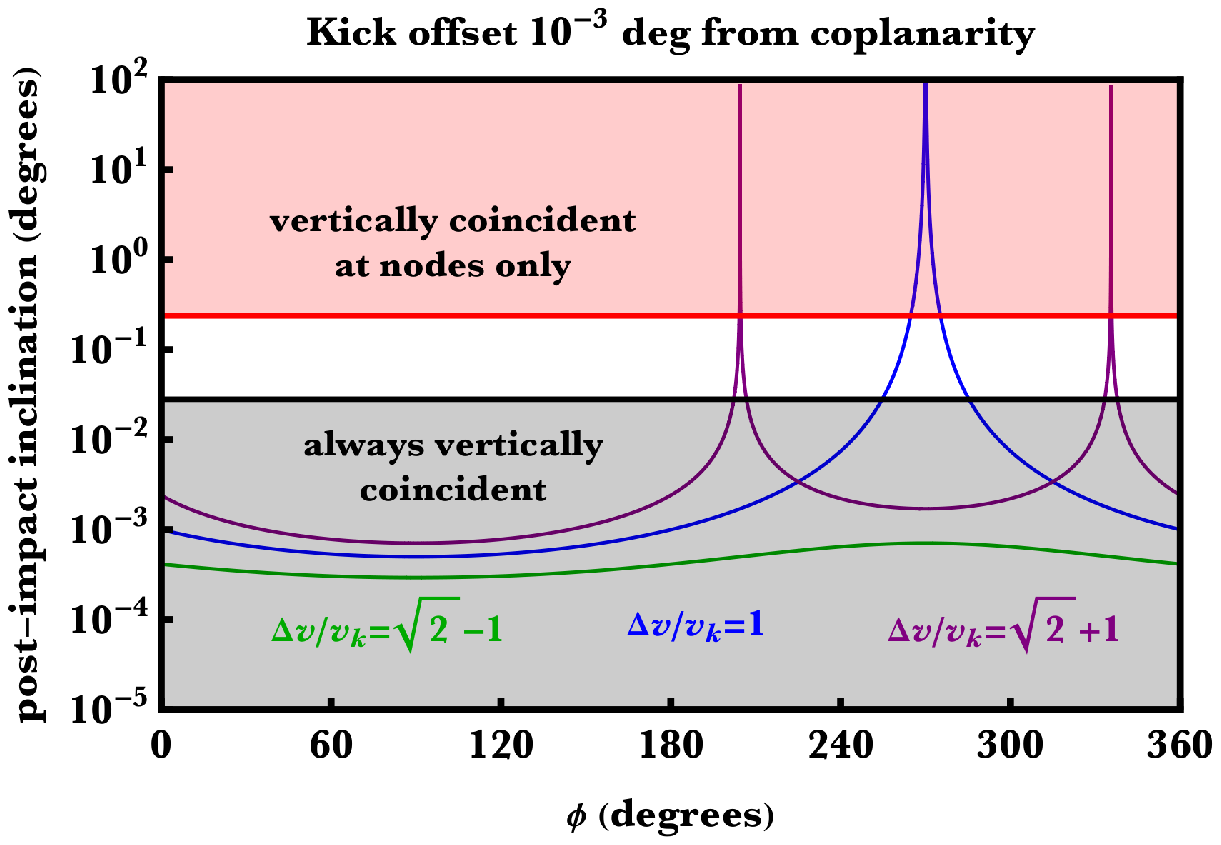}
 \includegraphics[width=9cm]{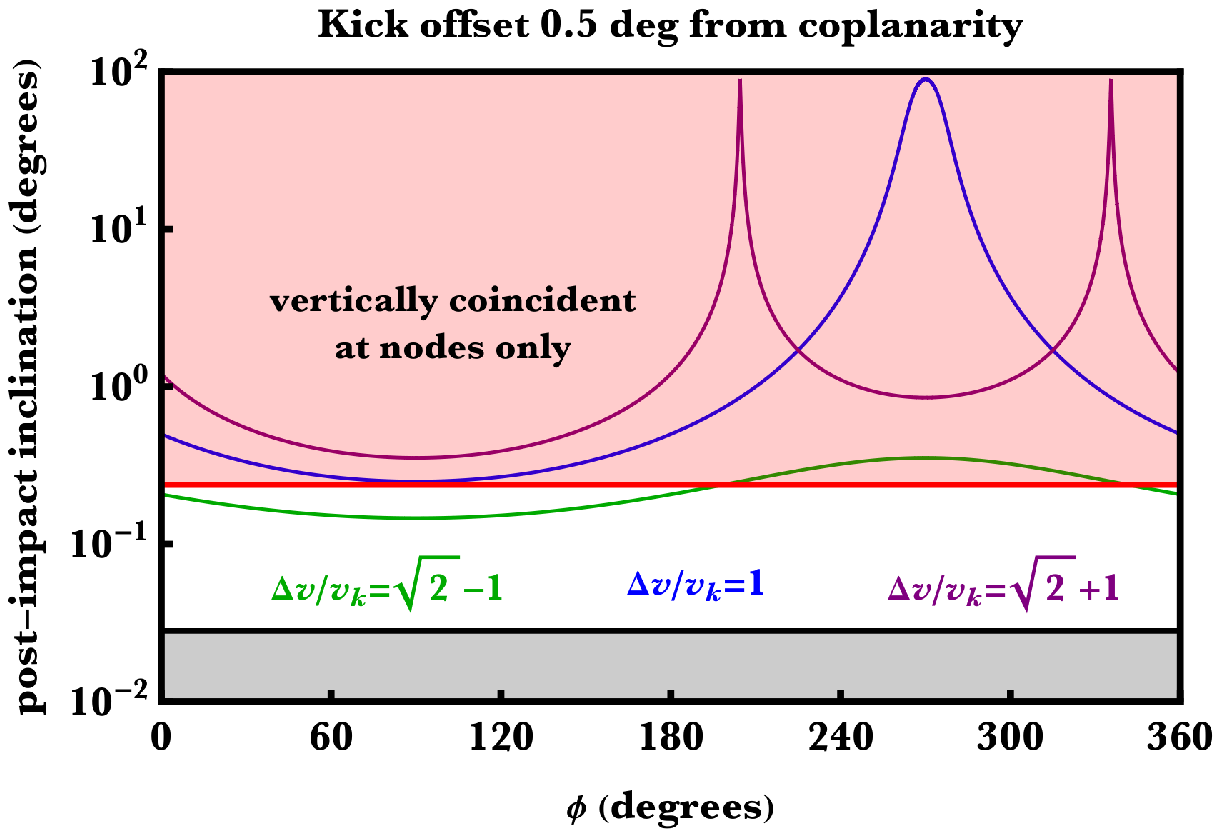}
\caption{
How the kick direction affects the inclination of the ejecta orbit. By assuming coplanarity amongst all TRAPPIST-1 planets, we plot, for three different values of the ratio of kick velocity to circular Keplerian velocity ($\Delta v/v_{\rm k}$), the dependence on the kick direction in the source's orbital plane $\phi$. The gray region corresponds to where the resulting ejecta orbital inclination is never large enough to exceed the radius of any TRAPPIST-1 target at any point in the orbit, and the red region corresponds to the opposite extreme, where vertical coincidence occurs only near the orbital nodes.
}
\label{incplots}
\end{figure*}

\subsection{Ejecta orbit characteristics}

The orbit of the ejecta, whose elements are denoted by primes, are related to the unprimed 
quantities (which refer to the source planet), through equations (\ref{aaprime})-(\ref{apo}) \citep{jacetal2014}.

Because known compact multi-planet systems are dynamically ``cold'' -- exhibiting low eccentricities -- henceforth we make the approximation that all planets are on circular orbits. Doing so greatly simplifies the analysis. For example, the upper limits for the planetary eccentricities in the TRAPPIST-1 system are all under 0.085. Further, because for any compact multi-planet system with habitable plants, we cannot yet know if panspermia has occurred, or will occur, at a particular time, we assume, without loss of generality, that the true anomaly $f=0$.  We hence constrain only when ejecta intersects the orbit of the target, and use equations (\ref{peri}) and (\ref{apo}) for this purpose. 

\subsection{Coplanarity restrictions}

First however, consider that in order for ejecta to hit the target, the ejecta must coincide with the target in all three spatial dimensions. This intersection is most easily achieved if the orbital planes of the source, ejecta and target lie close to one another as we can then be guaranteed of coincidence in one of the three dimensions. Hence, in this section, we quantify how coplanar the orbits of the ejecta, source and target must be to achieve spatial coincidence in the vertical direction (direction of the angular momentum vector of the source) throughout the debris orbit.

This condition is mathematically equivalent to $q' \sin{I'} \lesssim R$ or $Q' \sin{I'} \lesssim R$, when assuming that $R$ denotes planet radius, $q'$ denotes orbital pericentre, $Q'$ denotes orbital apocentre, and the reference plane is the one which connects a coplanar source and target. Note that unless the longitudes of ascending node can be measured, the mutual inclinations of all source-target pairs in a particular system will remain unknown\footnote{In the TRAPPIST-1 system, the longitudes of ascending node are so far unknown, and the maximum difference in measured inclinations is $0.21^{\circ}$, when neglecting errors.}. The approximations in these relations result from effects not considered here such as gravitational focusing and atmospheric drag.

If the kick direction is perpendicular to the source's angular momentum direction ($\theta = \pi/2$), then the ejecta orbit will be coplanar with the reference orbit ($I' = 0$). If, however, the kick direction deviates from these values, then the result is less obvious and is dependent on $\phi$ (and $e$ when not assumed to be zero). 

Figure \ref{incplots} illustrates the resulting dependence of $I'$ on $\phi$ for two different values of $\theta$, one where $\theta$ deviates from coplanarity by $10^{-3}$ degrees (left panel) and the other where the deviation is 0.5 degrees (right panel). 

In order to provide some context, we superimpose some results from the TRAPPIST-1 system on this figure. The gray region corresponds to where in the TRAPPIST-1 system $I' \le \arcsin{(\min{(a/R)})} = 0.029^{\circ}$ and the red region to where $I' \ge \arcsin{(\max{(a/R)})} = 0.24^{\circ}$. For this red region, the ejecta and target would be vertically coincident only near the nodes of their orbits, whereas for the gray region, they would be vertically coincident throughout their orbits (increasing the chances of impact). The purple curve, corresponding to $\Delta v/v_{\rm k} = \sqrt{2} + 1$, is a limiting case for which the ejecta may remain bound. As $\Delta v/v_{\rm k}$ is lowered and approaches zero, the resulting curves would be lower than the green curve. For a kick deviation offset from coplanarity exceeding about $0.86^{\circ}$, we find that all three curves lie entirely within the red region.

\subsection{Other spatial restrictions}

Having linked the angles of impact with the inclinations, we now turn to the other two spatial dimensions. The subsequent analysis greatly benefits from three simplifications, which are sufficient for this study. The first two are our continued assumption of circular and coplanar orbits. The third is that we consider the source and target planets in pairs, ignoring the influence of any other planets including those whose orbits lie in-between the pairs. This last assumption degrades in accuracy as the distance between the planets increases, because the debris will take longer to traverse this distance, and hence be diverted to a greater extent by extra bodies. However, these effects are sensitively dependent on the number, masses and locations of other planets in the system.

\subsubsection{Intersecting orbits}

In order for a collision to occur, the debris must be ejected onto an orbit which intersects the orbit of another planet. We can place bounds on this geometry by considering the pericentre and apocentre. Hence we begin by expressing these quantities as a function of the impacted planet's semimajor axis, the velocity ratio and $\phi$ (equations \ref{origsmallq}-\ref{chieq}). We wish to find the range of velocity kicks for which these pericentres or apocentres are achieved. Hence, inverting the equations yields, for inward motion towards the orbital pericentre,

\begin{equation}
\left( \frac{\Delta v}{v_{\rm k}} \right)_{\rm in} = \left\{
\begin{array}{ll}
  
  \frac{\left(a-q'\right)\left[\left(a+q'\right)\sin{\phi}+\sqrt{\xi}\right]}{q'^2-a^2\sin^2{\phi}},
  & \quad \sin^2{\phi} < \left(\frac{q'}{a}\right)^2 \\
  
  \frac{\left(a-q'\right)\left[\left(a+q'\right)\sin{\phi}-\sqrt{\xi}\right]}{q'^2-a^2\sin^2{\phi}},
  & \quad \sin^2{\phi} > \left(\frac{q'}{a}\right)^2

\end{array}
\right.
\label{FirstPiece}
\end{equation}

\noindent{}where

\begin{equation}
  \xi = q' \left[q' + \left(2a + q'\right) \sin^2{\phi} \right]
  .
\end{equation}

\noindent{}Recall that $a$ and $q'$ refer respectively to the source planet's semimajor axis and
the ejecta's orbital pericentre, and $\phi$ is the angle between the projection of the kick
direction onto the source's orbital plane and the star-source pericentre line.

For panspermia outward from the star, no piecewise function is
necessary, as in the following equation the denominator is always positive
and the term in square brackets is always negative.

\begin{equation}
\left( \frac{\Delta v}{v_{\rm k}} \right)_{\rm out} =  
  \frac{\left(a-Q'\right)\left[\left(a+Q'\right)\sin{\phi}-\sqrt{\zeta}\right]}{Q'^2-a^2\sin^2{\phi}}
\label{SecondPiece}
\end{equation}

\noindent{}where

\begin{equation}
  \zeta = Q' \left[Q' + \left(2a + Q'\right) \sin^2{\phi} \right]
  .
  \label{ThirdPiece}
\end{equation}

\noindent{}Equations (\ref{FirstPiece})-(\ref{ThirdPiece})
provide the necessary and sufficient conditions for kick speeds to propel ejecta into the orbit of another planet. 

\subsubsection{Application to TRAPPIST-1}

\begin{figure}
 \includegraphics[width=9cm]{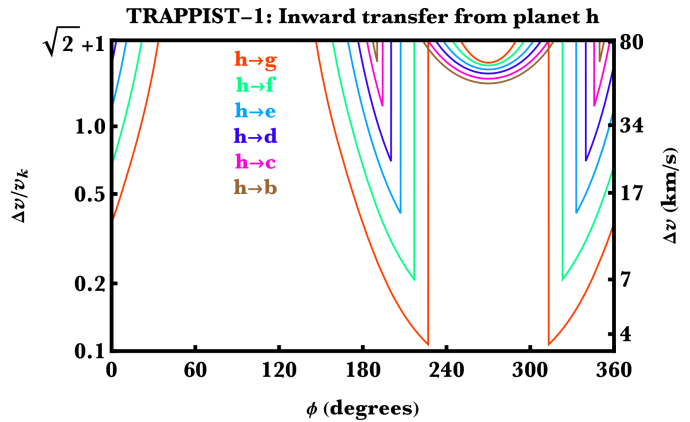}
\caption{
Inward panspermia in the TRAPPIST-1 system from planet h. Shown is the minimum speed kick required to place ejecta from planet h into the orbit of another planet, as a function of kick direction ($\phi$). For many values of $\phi$, the minimum kick speed exceeds the system escape speed (top axis of plot).
}
\label{perisample}
\end{figure}

In order to demonstrate how these equations can be applied to a real
system, Fig. \ref{perisample} illustrates the application of equation
(\ref{FirstPiece}) to the TRAPPIST-1 system. The figure displays inward panspermia from
planet h, as well as what minimum kick speeds are necessary to
thrust the ejecta just into the orbits of planets g, f, e, d, c and b.
Note that this value is highly dependent on $\phi$, and exceeds the
system escape speed for many values of $\phi$. Further, the dependence on
$\phi$ is non-monotonic.

\begin{figure}
  \vspace{1em}
  \includegraphics[width=9cm]{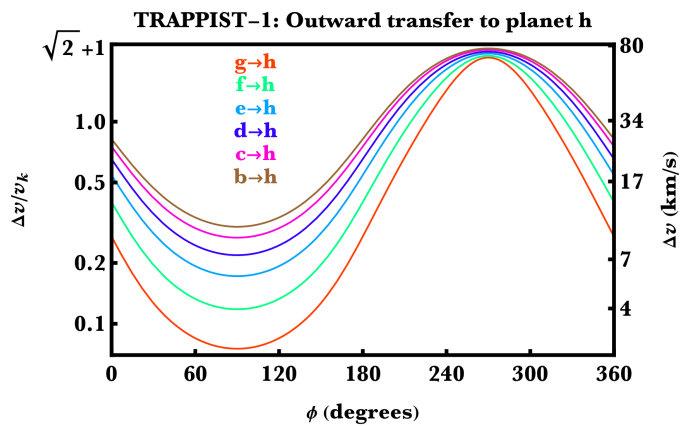}
  \caption{
Outward panspermia in the TRAPPIST-1 system to planet h. Shown is the minimum speed kick required to place ejecta from a planet into the orbit of planet h, as a function of kick direction ($\phi$). In all cases, the minimum kick speed is less than the system escape speed (top axis of plot).
}
\label{aposample}
\end{figure}

When we instead consider outward panspermia to planet h, the
functional form changes. Application of equation (\ref{SecondPiece})
yields Fig. \ref{aposample}. Here, for all values of $\phi$,
the minimum kick speed required to propel the ejecta to another planet's
orbit is smaller than the system escape speed. The smallest kick would
be from Planet h's nearest neighbour (planet g), whereas the greatest
kick is required from the planet furthest away (planet b).

Both Figs. \ref{perisample} and \ref{aposample} do not take into account the escape speed from the planets themselves. The escape speed of the TRAPPIST-1 planets are, from planet b to h moving outward, $\left\lbrace 9.90, 12.79, 8.15, 9.19, 9.02, 12.2, 9.35 \right\rbrace$ km/s. The relative velocity of the ejecta after escape from the planet is likely to be comparable to the escape velocity of the planet.

\subsubsection{When keeping $\phi$ fixed}

\begin{figure}
  \vspace{1em}
  \includegraphics[width=9cm]{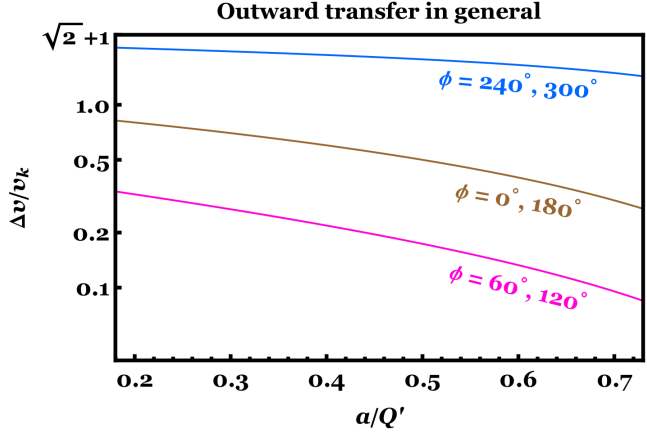}
  \caption{
Outward panspermia as a function of $\left(a/Q'\right)$ when $\phi$ is kept fixed. The $x$-axis includes the entire relevant range in the TRAPPIST-1 system; the upper end also roughly corresponds to outward panspermia from Earth's orbit to Mars' (assuming that they are on circular, coplanar orbits).
}
\label{nondim}
\end{figure}

If instead one has reason to assume a particular fixed value of $\phi$, then equations (\ref{FirstPiece})-(\ref{ThirdPiece}) may be considered as functions of $\left(q'/a\right)$ or $\left(a/Q'\right)$. As an example, the result for outward panspermia is shown in Fig. \ref{nondim} for six values of $\phi$. This plot may be applied to any planetary system, including the solar system. For example, outward panspermia from Earth to Mars \citep{woretal2013} corresponds roughly to the right axis of the plot. The escape speeds of those planets, however, differ by over an order of magnitude from those of the TRAPPIST-1 planets.

\subsubsection{Curve extrema}

Returning to Figs. \ref{perisample}-\ref{aposample}, in order to find the minimum value of these curves as a function of $\phi$, we consider all possible extrema of equation (\ref{FirstPiece}), which are found in equation (\ref{curveex}). For outward panspermia,

\begin{equation}
\max{\left(\frac{\Delta v}{v_{\rm k}}\right)_{\rm out}^{\rm abs}} = \sqrt{\frac{2Q'}{a + Q'} } + 1
\label{useful1}
\end{equation}

\noindent{}and $(\phi)_{\rm ex} = \pi/2$, which gives the absolute minimum

\begin{equation}
\min{\left(\frac{\Delta v}{v_{\rm k}}\right)_{\rm out}^{\rm abs}} = \sqrt{\frac{2Q'}{a + Q'} } - 1.
\label{useful2}
\end{equation}

\noindent{}These equations explain why the system escape speed corresponding to $\Delta v / v_{\rm k}$ ($= \sqrt{2} + 1$) is never reached. Because the edge of the system is beyond planet h, the ejecta will reach planet h before the edge of the system.

Alternatively, for inward panspermia, there are multiple real solutions: 
$(\phi)_{\rm ex} = \left\lbrace -\pi/2, \ \ -\arccos{\left[\pm \sqrt{a^2 - q'^2}/a\right]} \right\rbrace$.
Combined with the escape boundary, we have

\begin{equation}
\max{\left(\frac{\Delta v}{v_{\rm k}}\right)_{\rm in}^{\rm abs}} = \sqrt{2} + 1
,
\end{equation}

\begin{equation}
\min{\left(\frac{\Delta v}{v_{\rm k}}\right)_{\rm in,1}} = \sqrt{\frac{2q'}{a + q'} } + 1
,
\label{loc}
\end{equation}

\begin{equation}
\min{\left(\frac{\Delta v}{v_{\rm k}}\right)_{\rm in,2}} = \frac{a \left(a - q'\right) }{2q' \left(a + q'\right)}
.
\label{useful5}
\end{equation}

The absolute minimum could arise from either equations (\ref{loc}) or (\ref{useful5}). Equation (\ref{loc})
is the absolute minimum when $q'/a < (1/2)(\sqrt{2} - 1) \approx 0.21$; otherwise equation (\ref{useful5})
gives the absolute minimum. Equations (\ref{useful1})-(\ref{useful5}) may be useful constraints for any 
compact multi-planet system. 

\subsubsection{Probability distributions}

Given these constraints, we can now construct probability distribution functions for debris reaching the target planet's orbit . We 
obtain a probability $P$ of a given ratio ($\Delta v / v_{\rm k}$) being sufficiently high for the 
ejecta to reach the orbit of another planet as an explicit function of ($\Delta v / v_{\rm k}$), $a$ and either $q'$ or $Q'$,
by assuming some distribution for $\phi$. Because the cratering record on Solar system bodies indicates 
that ejecta are effectively isotropically distributed, we assume that a uniform distribution of $\phi$ holds
generally for other planetary systems. Consequently, the formulae we obtain may be applied widely 
if extrasolar systems experience similar cratering processes. Our final results are presented
in equations (\ref{out1})-(\ref{In4}).

\begin{figure}
  \vspace{1em}
 \includegraphics[width=9cm]{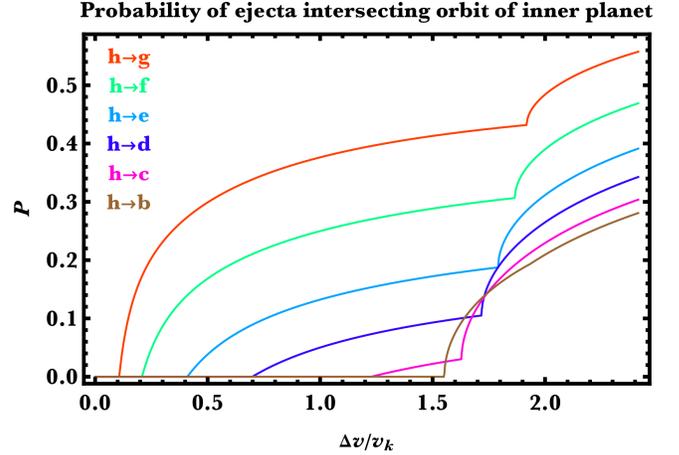}
\caption{
Application of the general algorithm to yield transfer probabilities (equations \ref{In1}-\ref{In4}) to the TRAPPIST-1 system. Shown is inward panspermia in to planet h. The only variables needed to generate this plot were the semimajor axes of the planets.
}
\label{probin}
\end{figure}

\begin{figure}
  \vspace{1em}
 \includegraphics[width=9cm]{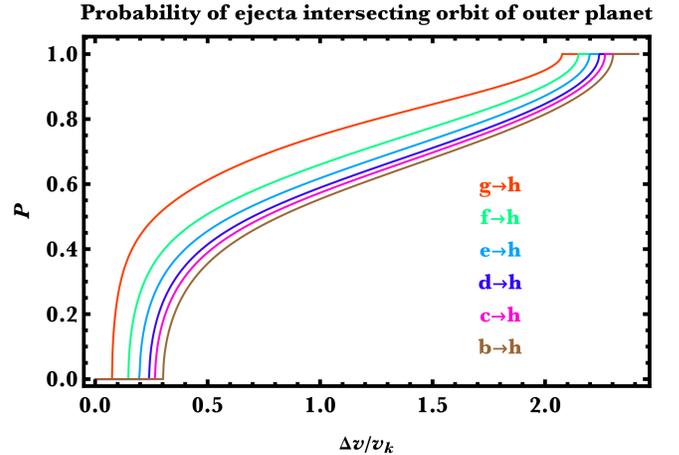}
\caption{
Similar to Fig. \ref{probin}, but for outward transfer of debris to the orbit of planet h (equations \ref{out1}-\ref{out3}). For high-enough kick velocities, ejecta will always have the capability of colliding with an outer planet.
}
\label{probout}
\end{figure}

Now we provide an example using these probability functions. We utilize the TRAPPIST-1 system
both in order to be consistent with the previous applications and because, fortuitously,
planets b and h
are sufficiently well-separated to allow us to sample a special case where $q'/a < (1/2)(\sqrt{2} - 1)$.
Recall that the only variables required to construct these functions are the semimajor axes of the
planets and the velocity ratio.

Figures \ref{probin} and \ref{probout} are the result. Note first that for inward panspermia, $P$ never
reaches unity, unlike for outward panspermia: the geometry responsible for
these relations are folded into
the curves. Second, for both directions, transferring to neighbouring planets is easier than for 
those further away. Third, the kinks in Fig. \ref{probin} ultimately result from the piecewise
nature of the velocity ratio in equation (\ref{FirstPiece}). The kink in the bottommost
curve occurs at $P=0$ because the separation of planets h and planet b is wide enough to cross
the critical threshold mentioned in the last paragraph.

\section{Impacting the target in one pass}

In the last section, we placed bounds on orbital properties of ejecta that could hit the target planet. In order to better quantify the probability of actually impacting the target, we now consider the location of the debris in space, rather than just its orbit.

Assuming that the ejecta orbit intersects or almost intersects with the orbit of the target, and that these orbits remain fixed, then expressions exist for the probability of collision in one pass. These expressions, pioneered by \cite{opik1951} and \cite{wetherill1967}, have led to substantial and wide-ranging applications. A recent updated and simplified series of derivations was provided by \cite{jeomal2017}. They found in their equations 29 and 37 the probability of collision per revolution of the ejecta, $\mathcal{P}$, as a function of (i) the orbital periods of the target and ejecta, (ii) the gravitational acceleration due to the parent star (assumed to be constant in the vicinity of the collision), (iii) the collision radius, (iv) the velocities of both objects at collision, and (v) the angle, $\lambda$, between the common direction of the velocity vectors and the vector from the star to the collision point.

We now consider these dependencies in more detail. Because we assume that the target is on a circular orbit, the gravitational acceleration at the collision point is $GM_{\rm star}/a''^2$ (recall that the target's orbital parameters and mass are denoted with double primes). The collision radius is the sum of the radii of the ejecta and target multiplied by the gravitational focusing factor $(1+ v_{\rm esc}''^2/|v'-v''|^2)$. These velocities can be expressed in orbital elements as $v'' = n''a''$ and $v' = n'a'\sqrt{(1+e'\cos{f'})(1-e'^2)}$ such that
$n'$ and $n''$ denote the mean motion of the ejecta and target. In this last expression, we can further relate $f'$ to $\phi$. For the circular source orbit case, and assuming coplanarity amongst the source, debris and target, equations (\ref{sinf}) and (\ref{cosf}) reduce to

\begin{equation}
  \tan{f'} = \frac{\left|1 + \left(\frac{\Delta v}{v_{\rm k}}\right) \sin{\phi} \right| \cot{\phi}}
             {2 + \left(\frac{\Delta v}{v_{\rm k}}\right) \sin{\phi} }
\end{equation}

\begin{figure}
 \vspace{2em}
 \includegraphics[width=9cm]{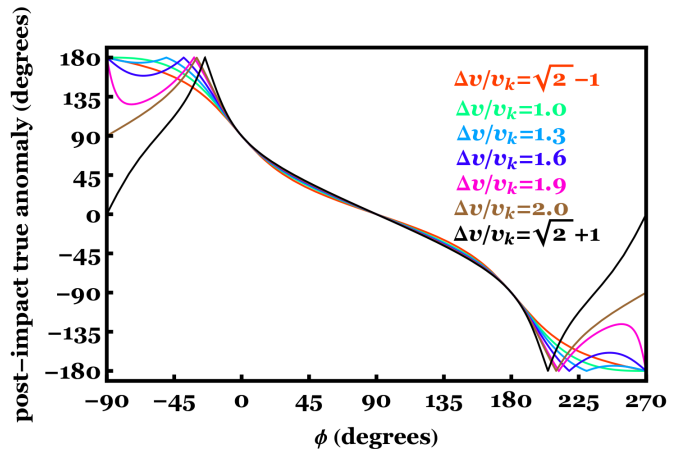}
\caption{
Relating $f'$ to $\phi$ for different kick speeds. 
}
\label{fPlot}
\end{figure}

\noindent{}which leads to Fig. \ref{fPlot}.

Adding these various components above together leads to the following expression for the collision probability per time in our formalism. Assuming that the target and ejecta are much less massive than the parent star allows us to concisely write

\[
  \mathcal{P} = \frac{c}{4\pi^2}
  \sqrt{
    G\left(\frac{M_{\rm star}^2}{M''}\right)
    \left(\frac{R'+R''}{a'^3a''}\right)
  }
\]
  
\begin{equation}
  \ \ \ \ \ \times \sqrt{\frac{\left|v' - v''\right|}{v'+v''}
  \left(1 + \frac{v_{\rm esc}''^2}{\left|v'-v''\right|^2} \right)
  \csc{\lambda}
  }
  \label{collprob}
\end{equation}

\noindent{}where $c$ is a constant equal to either $2\sqrt{2}$ for exactly intersecting orbits, or 1.7 for non-intersecting orbits. Equation (\ref{collprob}) accounts for both inward and outward panspermia because of the first absolute value. As emphasized by \cite{jeomal2017}, equation (\ref{collprob}) should not be used to predict a specific impact event in the past or future, but rather be used in a statistical way with streams of debris.

The time taken for the debris to travel from the source to the target planet -- a measure which influences the survival rates of microbes -- may be approximated by the inverse of equation (\ref{collprob}). This transit time is hence strongly influenced by $\left|v' - v''\right|$, which is the difference in speed of the ejecta and target planet. When these values are comparable, the inverse of equation (\ref{collprob}) becomes singular. Further, the angle $\lambda$ is crucial, and can reduce the transit time almost arbitrarily. These factors primarily explain the several order-of-magnitude spread in transit times in fig. 3 of \cite{krietal2017} for the TRAPPIST-1 system. Nevertheless, when comparing transit times in TRAPPIST-1 to other systems, the functional dependencies in equation (\ref{collprob}) may be useful. For example, the transit time scales inversely with the mass of the star and as the square root of the distance from the target planet.

\section{Ejecta characteristics}

Alternatively, if we assume that all of the ejecta represents a single boulder or pebble of mass $m'$, then we can speculate on the characteristics of this piece of ejecta, and whether it could be life-bearing. In this section, we neglect more complex possibilities, such as indirect supplementary ejection from small source planets. Source planets the size of Ceres, for example, have low-enough escape velocities and surface pressures to be susceptible to glaciopanspermia \citep{houtkooper2011,linloe2017b}. We also neglect the portion of the atmosphere -- even for large source planets -- that will invariably be ejected along with the debris \citep{berera2017}. The debris can act as an agent to transfer the source planet's atmospheric constituents to the target. If the transfer of material is extensive enough, as during perhaps a period of heavy bombardment \citep{deNetal2012,botnor2017}, then the atmosphere of the target planet may become less or more susceptible towards hosting life on its surface.

\subsection{Escaping the atmosphere}

First, consider that the radius and density of this piece of ejecta are related to $m'$ through $m' = (4\pi/3)\rho'R'^3$. This radius must be larger than the critical radius needed to maintain escape velocity from the source. If the source planet contains an atmosphere with surface pressure $p$ and gravitational acceleration $g$, then the radius $R'$ of a piece of ejecta which could escape an atmosphere is \citep{artiva2004}

\begin{equation}
R' \ge \frac{3p}{8g\rho}
\left[
\frac{\Delta v + v_{\rm esc}}
     {\Delta v - v_{\rm esc}}
\right], \ \ \ \ \ {\Delta v > v_{\rm esc}}
\label{ejecrit}
\end{equation}

\noindent{}where $g = GM/R^2$ such that any ejector could escape from an atmosphere-less source as long as $\Delta v$ is sufficiently high. Consequently, the minimum single-body ejecta mass which eventually hits the target and could initially escape from the source is

\begin{equation}
m' \ge \frac{9\pi}{128} 
       \left( \frac{\rho' p^3}{\rho^3 g^3} \right)
       \left[ 
\frac{\Delta v + v_{\rm esc}}
     {\Delta v - v_{\rm esc}}
       \right]^3, \ \ \ \ \ {\Delta v > v_{\rm esc}}.
\label{minmass}
\end{equation}

\begin{figure}
  \vspace{1em}
  \includegraphics[width=9cm]{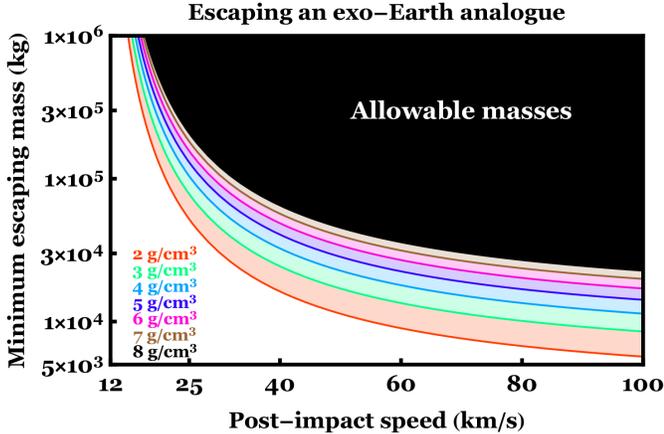}
\caption{
The minimum mass of an impact fragment that could be ejected from an Earth-analogue. The surface
pressure ($P$), surface gravity ($g$) and density of the Earth ($\rho$) were assumed. Each curve represents a different
impact fragment density ($\rho'$). The $x$-axis represents the post-impact speed $\Delta v$ is 
bounded from below by the escape speed. The plot demonstrates that near the escape speed, the minimum
mass needed to escape can vary by orders of magnitude. Otherwise, most post-impact
fragments with approximately at least the same mass as Big Ben (the bell) or the Hubble Space Telescope 
can escape an exo-Earth.
}
\label{MinEj}
\end{figure}

We plot equation (\ref{minmass}) in Fig. \ref{MinEj} assuming that the source planet is an
Earth-analog. The plot illustrates that atmospheric drag has the strongest effect when the
post-impact speed is close to the escape speed, and flattens out as $\Delta v$ increases
(note that for large $\Delta v$, equation \ref{minmass} becomes independent of $\Delta v$).

\subsection{Largest fragment}

Recall that $m'$ represents a fragment of the impact. Now we consider how to compute the largest possible size of an impact fragment, with radius $R'_{\rm max}$.
This deduction is based on detailed physics (porosity, spalling, Grady-Kipp processes, simple versus complex
crater formation) that are beyond the scope of this study. Further, certain dependencies which might
work for icy Solar system satellites \citep{bieetal2012,sinetal2013,alvetal2017} might not apply uniformly 
to all planets in extrasolar systems.

We just use \citep{miletal2000}

\begin{equation}
\frac{R'_{\rm max}}{R_{\rm i}} = 
\left( \frac{3 + W}{2} \right)
\left[  
\frac{T}{\rho \left(\Delta v^{2/3}\right) \left(v_{\rm i}^{4/3}\right)}
\right]
\label{avgR}
\end{equation}

\noindent{}where $R_{\rm i}$ and $v_{\rm i}$ are the radius and impact speed of the impactor,
$T$ is the tension at fracture, and $W$ is the Weibulls modulus \citep{affetal2006}.
For context, as mentioned by \cite{miletal2000}, $T = 0.1 \times 10^9$ Pa and $W = 9.5$ for basalt, a type of igneous rock.

We may relate the pre-impact and post-impact speed from equation (\ref{avgR}) by assuming that the 
energy of the impact is deposited at a depth which is comparable to $R_{\rm i}$.
Then \citep{melosh1984,melosh1988}

\begin{equation}
\frac{\Delta v}{v_{\rm i}} \approx \left(\frac{R_{\rm i}}{d} \right)^{2.87}
\end{equation}

\noindent{}where $d > R_{\rm i}$ is the distance from the ejection point to the centre of energy deposition
\footnote{See \cite{houetal1983} and \cite{dobetal2010} for alternative formulations.}. Consequently,
the ejecta speed cannot be larger than the impactor speed. Otherwise, the speed ratio is largely
determined by the detailed physics of the impact. As we do not pursue such detail here, we
suffice to leave $\Delta v/v_{\rm i}$ as a free parameter ranging from 0 to 1. Then, we can 
express equation (\ref{avgR}) for basalt (adopting $\rho = 3$ g/cm$^3$) as  

\begin{equation}
\frac{R'_{\rm max}}{R_{\rm i}} = \left( \frac{v_{\rm i}}{0.456 \ \frac{\rm km}{\rm s}} \right)^{-2}  
                              \left( \frac{\Delta v}{v_{\rm i}} \right)^{-2/3}.
\label{Rratio}
\end{equation}

\begin{figure}
  \vspace{1em}
  \includegraphics[width=9cm]{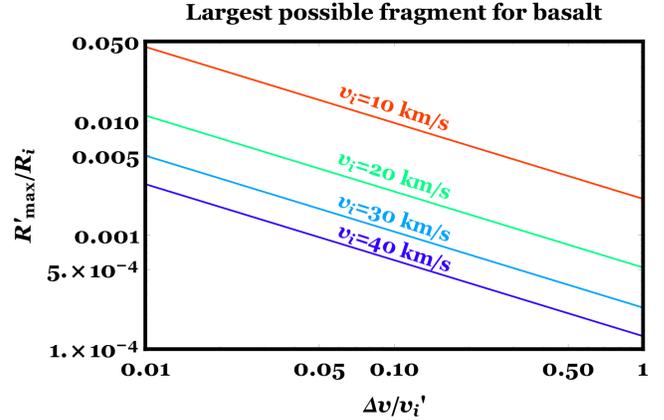}
\caption{
The maximum impact fragment size in terms of the initial radius of the impactor, for
different incoming impactor speeds. The tension
at fracture is assumed to be $0.1 \times 10^9$ Pa, and the density and Weibulls modulus are
consistent with basalt.
}
\label{Radmax0}
\end{figure}

\noindent{}We plot equation (\ref{Rratio}) in Fig. \ref{Radmax0}. The plot illustrates that,
in general, fragment radii are no more than about 5\% of the radius of their progenitor.

\subsection{Liberating material}

\begin{figure}
  \vspace{1em}
  \includegraphics[width=9cm]{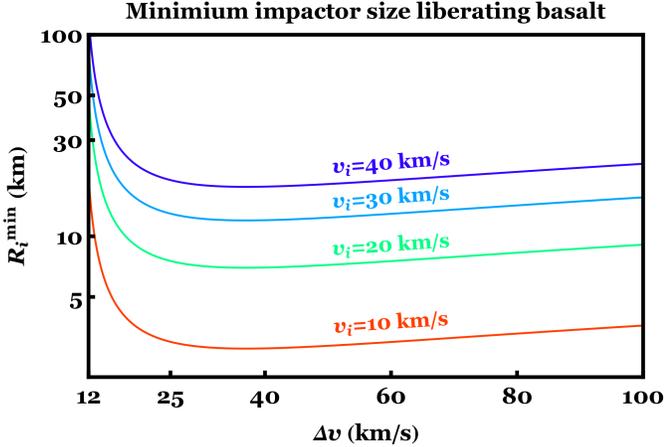}
\caption{
The minimum impactor size that could liberate material from the planet's gravitational well.
The impactor and planet have densities of 3 g/cm$^3$, which is consistent with basalt. 
The figure illustrates that panspermia could occur typically only if the impactor size is at least on
the order of tens of km in size.
}
\label{minlib}
\end{figure}

Now we may consider the minimum impactor size that could liberate material.  If $R' = R'_{\rm max}$ is taken
to be the minimum size of ejecta that can escape the atmosphere (through equation \ref{ejecrit}), then 
it follows that the corresponding $R_{\rm i} = R_{\rm i}^{\rm min}$ (through equation \ref{Rratio})
must be the minimum size impactor that can liberate material. Combining these equations, and assuming
basalt for the impactor and an Earth-sized, Earth-mass planet, yields
Fig. \ref{minlib}. The figure demonstrates that generally an impactor must have a radius of at least
tens of km in order to be the catalyst for panspermia.

\subsection{Destroying the source}

Finally, if the impactor is too energetic, then it could sterilize, eviscerate or
break apart the source planet. Any of these processes would inhibit prospects for panspermia.
Here, we do not analyze the consequences in detail, but merely provide bounds for the
impactor in the most extreme case of breakup.

We can place an upper bound on the
maximum size of the impactor by considering the maximum specific energy $E_{\rm max}$ imparted to a source for which the source will remain intact. The conditions for catastrophic disruption have 
an extensive associated literature, and can be characterized through a variety of metrics.
For example, see \cite{benasp1999} and the hundreds of more recent papers which cite that one.

We adopt the explicit formalism in Section 5 of \cite{movetal2016}. They show that
the source planet would break apart if the following condition is met:

\begin{equation}
\left(\frac{1}{2}\right)\left(\frac{\epsilon M + M_{\rm i}}{M + M_{\rm i}}\right)
                        \left(\frac{M M_{\rm i}}{M + M_{\rm i}}  \right) v_{\rm i}^2
> E_{\rm max}
\label{colcon}
\end{equation}

\noindent{where}

\begin{equation}
\epsilon \equiv \left\{
\begin{array}{ll}
  
  \frac{3 R_{\rm i} l^2 - l^3}{4 R_{\rm i}^3},
  & \quad l < 2 R_{\rm i} \\
  
  1,
  & \quad l \ge 2 R_{\rm i}
\end{array}
\right.
,
\end{equation}

\begin{equation}
l \equiv \left(R + R_{\rm i} \right) \left(1 - \sin{\theta} \right)
,
\end{equation}

\begin{equation}
E_{\rm max} = \left(13.4 \pm 10.8\right)
\left[\frac{3GM^2}{5R} + \frac{3GM_{\rm i}^2}{5R_{\rm i}} + \frac{GMM_{\rm i}}{R + R_{\rm i}} \right]
\label{Emax}
\end{equation}

\noindent{}with $M_{\rm i}$ being the mass of the impactor, and $\theta$ the impact angle, such that 
a head-on collision corresponds to $\theta=0$.
The numerical range given in equation (\ref{Emax}) applies for $0 \le \theta \lesssim 45^{\circ}$.

We can quantify equations (\ref{colcon})-(\ref{Emax}) in a simplistic but comprehensive manner
by reducing the number of degrees of freedom in the equation. Assume that the collision is head-on
and the impactor and source planet are made of the same substance (or more technically have equal
densities). Then we can reduce the condition to a function of two ratios as

\[
\left( \frac{v_{\rm i}}{v_{\rm esc}} \right)^2
>
\left(1.10 \pm 0.58 \right)
\]

\begin{equation}
\ \ \  
\times \left[
8 + 3 \left(\frac{R_{\rm i}}{R} \right)^{-3}
  - 5 \left(\frac{R_{\rm i}}{R} \right)
  + 8 \left(\frac{R_{\rm i}}{R} \right)^2
  + 3 \left(\frac{R_{\rm i}}{R} \right)^5
\right]
.
\label{vandr}
\end{equation}

\begin{figure}
  \vspace{1em}
  \includegraphics[width=9cm]{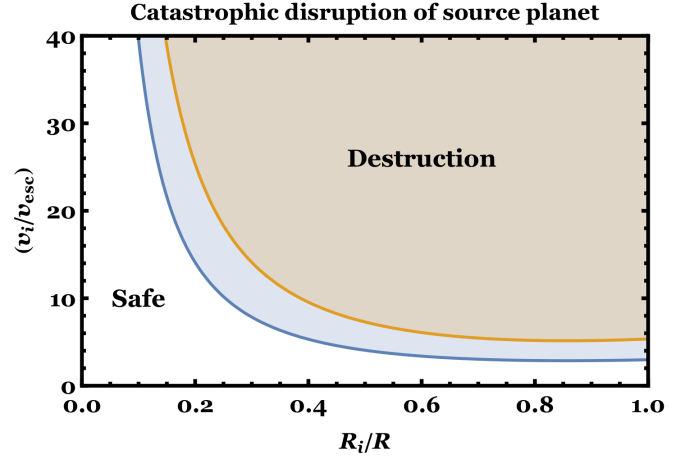}
\caption{
The speed and radius of an impactor that would catastrophically destroy the source planet. The collision
is assumed to be head-on, and the impactor and source planet are assumed to be made of the same material.
The two curves represent the bounds of the model prediction from the numerical coefficient of equation (\ref{vandr}).
If the impactor radius is less than about 10\% of the source planet's, then the required impact speed for
a destructive collision would be unrealistically high.
}
\label{Radmax}
\end{figure}

\noindent{}Figure \ref{Radmax} illustrates the phase space of equation (\ref{vandr}). 
The plot demonstrates that for common definitions of asteroid and planet, an asteroid 
could never destroy a planet. However, the result of a Mars-sized object colliding 
with an Earth-sized object is less clear.

\subsection{Temperature at impact}

We now have some idea of the size of potential ejectors. Returning our attention to panspermia, we must also consider temperature and how it will affect the ability of life to survive such an impact. 
Fig.~4 of \cite{weihea2016} illustrates averaged ejecta temperatures as a function of both crater diameter and surface heat fluxes on Mars. 
They find a range of 215 - 600 K for crater diameters between 0 - 150 km in roughly linear relationships for surface heat fluxes of 20 - 100 mW/m$^2$. 
Therefore, the temperatures generated are a strong function of crater diameter. 

Despite this wide range in temperature, simulations have shown that a significant proportion of the ejecta will not 
exceed 100 $^\circ$C due to the existence of a `spall' zone \citep{melosh1984}. 
This zone comprises much of the surface layer undergoing the impact and refers to a region in which the shock 
wave from the impactor is effectively cancelled via superposition with its reflected counterpart. 
These relatively low-temperature fragments would offer more favourable conditions to any lifeforms residing upon them. 
The following section will discuss specific forms of life that have shown great promise when tested against the panspermia hypothesis, 
from initial ejection (Section 5.1) through interplanetary travel (Section 5.2) to eventual atmospheric entry upon 
reaching the target planet (Section 5.3).

\section{Survival of micro-organisms}

Over the past few decades, it has become possible to simulate the three stages of panspermia: (1) initial 
ejection from the impacted planet, (2) the subsequent journey through interplanetary space, and finally 
(3) impact with the target planet. 
Each step provides a new set of challenges to the survival of life.
This section aims to supply a brief account of how certain micro-organisms have fared when placed in 
environments that are reminiscent of the panspermia process, and places these constraints into an extrasolar
planetary context. A more comprehensive review of the near-Earth and Solar system contexts
can be found in \cite{horetal2010}.

\subsection{Planetary ejection}

A precondition for ejecta to be produced is the existence of impactors. Not every exoplanet host star contains compact asteroid belts \citep{marliv2013}. 
However, evidence from white dwarf planetary systems indicate that between one quarter and one half of Milky Way white dwarfs host asteroid belts
or Kuiper belt analogs \citep{koeetal2014} 
that are dynamically excited by a number of mechanisms involving planets \citep{veras2016}, the same fraction of main sequence stars thought
to host planets throughout the Galaxy \citep{casetal2012}\footnote{No exoplanets have so far been discovered in other galaxies, and therefore the prevalence
of impactors there is unconstrained, although habitability prospects may still be assessed \citep{staetal2017}.}. Further, the transfer of either biomolecules 
(pseudo-panspermia; \citealt*{linloe2017a}) or dead organisms (necropanspermia; \citealt*{wesson2010}) imposes less stringent requirements than the transfer of living organisms.

In compact extrasolar systems, we might expect billions of rocks to be transferred between their planets, as
it has been estimated that in the much more expansive solar system hundreds of millions of rocks could 
have already been ejected from the spall zones of Martian impacts and made their way to Earth \citep{miletal2000}. 
Such a healthy estimate has provided motivation for a number of investigators to conduct
simulations which attempt to investigate the survivability of an impact large enough to eject material from Mars. 
One study, by \cite{stoetal2007}, applied pressures of 5 - 50 GPa to micron-thin layers of different micro-organisms. 
This pressure range, applied via high explosives, is thought to be typical of Martian impact ejection \citep{artiva2004}. 
Spores of \textit{Bacilus subtilis} exhibited survival fractions ($N/N_0$ where $N$ is the number of surviving cells and $N_0$ is the original number of viable cells) of 10$^{-4}$ under a pressure of 42 GPa. 

Bacterial spores are resilient casings that contain identical genetic information to their corresponding micro-organism. Thought to be due to their low water content, the cores of bacterial spores have been found to exhibit notably low enzyme activity. 
This property is believed to contribute to the resilience of the spores, alongside the fact that the bacterial DNA is mixed with acid-soluble proteins, which aid in enzymatic reactivation (Setlow 1995). 
Besides the application of high explosives, the planetary ejection stage has also been simulated by firing projectiles at layers of spores, with similar survival rates to that outlined above \citep{buretal2004,fajetal2009}. 

\subsection{Journey through interplanetary space}

Following successful ejection from the impacted planet, the life must then endure the trip to the target planet. 
Although not as violent as impact-driven ejection, this stage of interplanetary panspermia has been shown to be equally as deadly for a number of micro-organisms, owing to the deleterious conditions present in the space environment. 
The space between planets is vast and empty; vacuum pressures can drop as low as 10$^{-14}$ Pa, inducing severe desiccation. Exposure to both stellar and Galactic cosmic radiation can also be highly damaging, with stellar UV posing the biggest threat. Temperature extremes during interplanetary transit have the potential to rival those of the ejection phase, depending on the orientation of the fragment's orbit around its host star. Together, these form a lethal concoction for many micro-organisms. Some, however, have exhibited impressive resilience against the harsh conditions of space.

The seven planets of TRAPPIST-1 orbit an ultra-cool M dwarf star. Due to the star's relatively low temperature, 
its habitable zone lies very close in (at hundredths of an au). As such, it is likely to play host to a radiation 
environment that is much more damaging than the one surrounding the Earth. Investigations into the X-ray/EUV irradiation 
of the planets were undertaken by \cite{wheetal2017}. They deduced that TRAPPIST-1 has an X-ray luminosity similar 
to that of the quiet Sun; it has been hypothesised that such a high flux at these wavelengths could have stripped 
the planets of their atmospheres \citep{donetal2017,roekan2017}, raising severe doubts regarding their habitability. The planetary atmospheres are also expected to alter frequently due to persistent flaring events \citep{videtal2017}, which can penetrate magnetospheres \citep{garetal2017} and also affect the surface-based biospheres \citep{caretal2018} in conjunction with geochemistry \citep{baretal2017}. It would appear, therefore, that the TRAPPIST-1 planets are unlikely to support atmospheres that would enable the harbouring of life.

For more Solar system-like exoplanetary systems, we can consider the
numerous exposure missions that have taken place in low Earth orbit (LEO), namely at altitudes less than 2000 km. 
It is important to note that the results of LEO experiments provide a mere estimate of the survivability of panspermia; 
such orbits are relatively close in proximity to the Earth, and therefore fail to accurately mirror the conditions 
expected during an interplanetary transit. For instance, the minimum vacuum pressure in LEO is approximately 
10$^{-6}$ -- 10$^{-7}$ Pa, several orders of magnitude higher than would be experienced for the majority of a 
planet-planet trip. Furthermore, the magnetic field of the Earth would shield the lifeform from much of the 
harmful cosmic radiation that would be plentiful in other regions of the Solar System. 
Nevertheless, LEO missions have contributed vastly to our understanding of the survival limits of copious micro-organisms, 
placing constraints on the plausibility of panspermia as a concept.

\subsubsection{Long Duration Exposure Facility}

Still holding the record for the longest exposure to LEO, NASA’s Long Duration Exposure Facility (LDEF) subjected spores of \textit{B. subtilis} to a combination of the space vacuum, solar UV and multiple components of Galactic cosmic radiation \citep{horetal1994}.
In accordance with several other astrobiological experiments, solar UV was found to cause most damage due to its tendency to target and break DNA strands within the spores. 
Prospects for survival were greatly improved with adequate shielding from the UV in place; around 70 \% of \textit{B. subtilis} spores were able to survive 6 years of exposure to the LEO vacuum when mixed with the sugar glucose. 

\subsubsection{EURECA}

Similar conclusions were drawn from the results of the EURECA mission, which reported a complete loss of 
viability of \textit{Deinococcus radiodurans} cells following 9 months of exposure to LEO \citep{dosetal1995}. 
Up to 12 DNA double strand breaks were observed per chromosome in samples exposed to the solar UV, although shielded 
cells also showed complete inactivation. 
These findings were surprising, as \textit{D. radiodurans} is known to be incredibly resistant to desiccation and 
radiation; the bacterium can withstand radiation doses of 5000 Gray (around 1000 times a typically lethal dose for humans) 
with no loss of viability \citep{mosetal1971}. 

A more recent study has found that \textit{D. radiodurans} cells can survive many more dessication-induced DNA double strand breaks than those observed in the EURECA mission, due to the fact that the genome is able to reassemble before each successive cycle of cell division \citep{coxbat2005}. 
As such, the survival rate observed following exposure to LEO conditions could have been higher than that inferred from the experimental results. 
Regardless, it is clear that the combination of stellar UV radiation and space vacuum form a deadly cocktail, survivable only for the most resilient of lifeforms known to inhabit our Earth.
Indeed, vacuum-induced dehydration has been found to alter DNA photochemistry in such a way as to enhance the UV sensitivity of \textit{B. subtilis} spores ten-fold in comparison to irradiation at atmospheric pressure \citep{horneck1998,nicholson2000}.

\subsubsection{Biopan}

Owing to these early findings, a general consensus has emerged that adequate shielding from the harmful environment of 
interplanetary space must be in place for micro-organisms, such as bacterial spores, lichens and tardigrades, to stand a chance of surviving panspermia.

\paragraph{Shielding}

\cite{miletal2000} provide a thorough investigation of shielding for the case of Earth and Mars,
 and \cite{cockno1999} provide a thorough summary of shielding mechanisms from UV radiation.
The effects of shielding were explored as part of the series of experiments that took place using the Biopan facilities 
aboard various Foton satellites \citep{horetal2001}. 
A survival fraction of 10$^{-6}$ was obtained when \textit{B. subtilis} spores were exposed to the full LEO environment, 
whilst much higher fractions of 0.5 - 0.97 were determined for shielded samples. 
Clay shielding was found to be ineffective when placed in the form of a `shadowing' layer; much more protection was 
received if the spores were mixed in with the clay, or ground meteorite powder. 
Importantly, the samples consisted of multilayers of spores; the outer layers would have encountered the full extent 
of LEO conditions, inactivating quickly and potentially forming a protective `crust', offering added protection to the 
innermost layers of spores. 
It is also thought that endolithic micro-organisms, residing in microcracks present within rocks, likely exist in the
 form of biofilms embedded within a complex matrix of sugar molecules \citep{cosetal1987}. 
This configuration would provide additional protection against the space vacuum. 
As such, lifeforms mixed within a layer of rock or clay are likely to receive much greater shielding from both the 
stellar UV and the space vacuum.

\paragraph{Lichens}

During the Biopan 5 mission, thalli of the lichens \textit{Xanthoria elegans} and \textit{Rhizocarpon geographicum} were exposed to the space vacuum and selected wavebands of the solar UV for 14.6 days \citep{sanetal2007}. A lichen comprises a stable symbiotic interaction between fungi and/or cyanobacteria. Lichens can be endolithic, growing between the grains inside rock, and are commonly found in mountainous regions. They have been found to survive complete water loss throughout periods of severe desiccation \citep{kraetal2008} and withstand higher than average levels of UV radiation. Following the exposure to LEO, 83 \% of \textit{X. elegans} cells were found to have intact membranes, whilst a similarly high survival rate of 71 \% was determined for \textit{R. geographicum}. 

Furthermore, full photosynthetic recovery was observed, even for samples exposed to over 99 \% of the solar light. The lichens contain certain pigments that provide screening from the UV, heightening protection during exposure, such as parietin phenolic acids \citep{soletal2003}. Similar UV-screening properties were exhibited by cells of the halophilic cyanobacterium \textit{Synechoccus} following two weeks of exposure to LEO as part of the Biopan 1 series of experiments \citep{manetal1998}.
Interestingly, \textit{X. elegans} has also been tested in simulations of the planetary ejection stage of panspermia. The lichen fared similarly to \textit{B. subtilis} spores, with survival rates dropping by only four orders of magnitude upon the application of 50 GPa pressure \citep{horetal2008}.

\paragraph{Tardigrades}

Biopan 6, on the other hand, provided the first testing ground for tardigrades in space \citep{jonetal2008}. Tardigrades have been identified as one of the most resilient animals on Earth, so are a natural choice for testing in LEO. They have been found to survive extreme temperatures and pressures for significant periods of time \citep{henetal2009,horetal2009}, and show incredible resistance to radiation, surviving doses of up to 5000 Gray \citep{hasetal2016}. A recent study by \cite{sloetal2017} deduced that tardigrades are likely to survive any mass extinction event with an astrophysical cause, such as a nearby supernova, gamma-ray burst or large asteroid impact. 

In a similar way to bacterial spores, tardigrades can undergo a process known as cryptobiosis, whereby metabolic processes shut down in a reversible fashion during times of extreme stress. One particular form of cryptobiosis, known as anhydrobiosis, is of particular relevance to our discussion of survival in space. In this process, a tardigrade will contract and lose the vast majority of its water content, enabling cell stabilisers like trehalose to be formed and metabolism to, in the most extreme cases, be temporarily halted altogether \citep{weletal2011}. Samples of the tardigrade species \textit{Milnesium tardigradum} and \textit{Richtersius coronifer} survived exposure to the LEO vacuum very well. Combined exposure to both the space vacuum and solar/Galactic radiation resulted in reduced, yet still finite, survival for both species tested. Tardigrades therefore have joined bacterial spores and lichens in the list of lifeforms that have survived exposure to the full LEO environment.

\subsubsection{EXPOSE}

The most recent results obtained from exposure missions in LEO are those of the European Space Agency's EXPOSE facilities, mounted aboard various modules of the International Space Station. Conducted upon EXPOSE-E, the LIFE experiment subjected a variety of eukaryotic organisms to long-term exposure (1.5 years) for the first time \citep{onoetal2012}. Most notably, \textit{X. elegans} once again achieved full photosynthetic recovery, provided the samples were shielded from UV irradiation. The AMINO experiment, which took place aboard the EXPOSE-R facility, exposed organic molecules to LEO both in their natural state and embedded in meteorite powder \citep{beretal2015}. Chosen for the key roles they play in the formation of macromolecules considered essential for life, the amino acids glycine, alanine and aspartic acid showed minimal deterioration following exposure, with 72 \% of glycine remaining unaffected in unshielded form. Samples of the prokaryote \textit{Halorubrum chaoviator}, a halophilic archaeon, were exposed to LEO as part of the OSMO experiment \citep{mancinelli2015}. If shielded from the solar UV, the archaea exhibited 90 \% survival rates. 

\subsection{Atmospheric entry}
From the many exposure experiments that have taken place in LEO, it is clear that the deleterious conditions in space can have a devastating effect on many micro-organisms. However, it is also clear that a number of lifeforms possess the necessary resilience to survive in such hazardous environments, especially when adequate shielding is in place. We now turn our attention to the final stage of material transfer: atmospheric entry upon reaching the target planet.

Because entry speeds range from 12 to 20 kms$^{-1}$ for typical asteroids, the overall process can occur in the space of a few tens of seconds \citep{nicholson2009}. Frictional heating over this rapid timescale leads to the formation of a fusion crust on the surface of the meteorite. This crust ensures that the heating fails to penetrate further than the first few millimetres of material, allowing the interior to maintain a relatively constant temperature throughout. Provided the target planet possesses an atmosphere, the eventual impact with the surface will occur at terminal velocity (50 ms$^{-1}$ for Earth), a far tamer value than what is involved in the planetary ejection phase. 

Thus far, the best attempts to assess the ability of micro-organisms to survive meteoric entry have been those of the STONE missions, conducted upon the recovery module heat shields of the same Foton satellites used to host the Biopan 5 and 6 facilities \citep{paretal2008,fouetal2010}. The entry speed was measured to be 7.7 kms$^{-1}$, falling short of the expected speeds for asteroids provided above. Nevertheless, none of the micro-organisms tested showed any signs of viability following retrieval, most notably \textit{B. subtilis}. For one of the samples, the fusion crust was found to be around 5 cm deep, possibly due to cracks in the surface of the shield. It would seem, therefore, that further experimentation is required to make any sort of conclusion regarding the survivability of the entry stage of panspermia.

\section{Conclusions}

The strong prospects for future discoveries of habitable multi-planet systems prompted us to analyze several aspects of panspermia and derive new results. Here, we have applied an impulse formalism from \cite{jacetal2014} to generate orbital constraints on life-bearing ejecta travelling between planets in the coplanar circular case (equations \ref{FirstPiece} and \ref{SecondPiece}). Resulting analytic probability distributions depend only on the semimajor axes of the source and target planets (equations \ref{out1}-\ref{In4}) and can be readily applied to compact multi-planet systems. We have also repackaged and consolidated physical relations that are associated with ejecta to fit within one framework (minimum radius and mass to escape atmosphere: equations \ref{ejecrit}-\ref{minmass}; largest impact fragment: equations \ref{avgR} and \ref{Rratio}; minimum impactor size to liberate material: Section 4.3; speed and impactor radius to destroy source: equation \ref{vandr}). We finally included biological constraints from impact, interplanetary travel and atmospheric entry (Section 5). We hope that our results will represent useful tools to analyze future discoveries of compact multi-planet habitable systems.

\section{Acknowledgments}

We thank both referees for particuarly helpful and specific comments on the manuscript, resulting in an improved document.
DV gratefully acknowledges the support of the STFC via an Ernest Rutherford Fellowship (grant ST/P003850/1), and has received 
funding from the European Research Council under the European Union's Seventh Framework Programme (FP/2007-2013)/ERC Grant 
Agreement n. 320964 (WDTracer). DJA is supported by STFC through consolidated grant ST/P000495/1, and JAB is supported 
through STFC grant ST/R505195/1. APJ acknowledges support from NASA grant NNX16AI31G.

\appendix

\section{Appendix A: Orbit transfers}

In this appendix we provide formulae referenced in Section 2 and whatever additional prescriptions are needed to derive results. First, the orbit of the ejecta are related to the source planet through \citep{jacetal2014}

\begin{equation}
\frac{a}{a'} = 1 - \left(\frac{\Delta v}{v_{\rm k}}\right)^2 
 - \frac{2}{\sqrt{1-e^2}} \left(\frac{\Delta v}{v_{\rm k}}\right) \sin{\theta} \left[ \sin{\left(\phi - f\right)} + e \sin{\phi} \right],
\label{aaprime}
\end{equation}

\begin{equation}
e'^2 = 1 - \left(1 - e^2\right) \left( H \frac{a }{a'} \right), 
\end{equation}

\begin{equation}
  \cos{I'} =
  \frac{1}{\sqrt{H}}
  \left[
    1 + \frac{\sqrt{1 - e^2}}{1 + e \cos{f}}
    \left(\frac{\Delta v}{v_{\rm k}} \right) \sin{\theta} \sin{\left(\phi - f\right)}
    \right],
\end{equation}

\begin{equation}
  \sin{f'} = \frac{\sqrt{H}}{e'}
  \left[e \sin{f} + \sqrt{1-e^2} \left( \frac{\Delta v}{v_{\rm k}} \right) \sin{\theta} \cos{\left(\phi - f\right)}  \right],
  \label{sinf}
\end{equation}

\noindent{}and

\begin{equation}
  \cos{f'} = \frac{1}{e'}
  \left[H \left(1 + e \cos{f} \right) - 1 \right],
  \label{cosf}
\end{equation}

\noindent{}where

\begin{equation}
H = 1 + 2\frac{\sqrt{1-e^2}}{1+e\cos{f}} \left( \frac{\Delta v}{v_{\rm k}} \right) \sin{\theta} \sin{\left[\phi - f\right]}
+ \frac{1 - e^2}{\left(1 + e \cos{f}\right)^2}
  \left( \frac{\Delta v}{v_{\rm k}} \right)^2
  \left[\cos^2{\theta} + \sin^2{\theta}\sin^2{\left(\phi - f\right)}  \right]
  ,
\end{equation}

\noindent{}and where $e'$ and $f'$ refer to the eccentricity and true anomaly of the ejecta orbit. We also define the pericentre and apocentre of the ejecta orbit as

\begin{equation}
q' = a' \left(1 - e' \right)
,
\label{peri}
\end{equation}

\begin{equation}
Q' = a' \left(1 + e' \right)
.
\label{apo}
\end{equation}

For coplanar and circular orbits, the pericenter ($q'$) and apocentre ($Q'$) of the debris' orbit are constrained by

\begin{equation}
q' \le a\left[ \frac{1 - \sqrt{\chi} \left(\frac{\Delta v}{v_{\rm k}}\right) }{1 - \left(\frac{\Delta v}{v_{\rm k}}\right)^2 - 2 \left(\frac{\Delta v}{v_{\rm k}} \right) \sin{\phi}}   \right]
,
\label{origsmallq}
\end{equation}

\begin{equation}
Q' \ge a\left[ \frac{1 + \sqrt{\chi} \left(\frac{\Delta v}{v_{\rm k}}\right) }{1 - \left(\frac{\Delta v}{v_{\rm k}}\right)^2 - 2 \left(\frac{\Delta v}{v_{\rm k}} \right) \sin{\phi}}   \right]
,
\label{origlargeq}
\end{equation}

\noindent{}where

\begin{equation}
  \chi = 
  1 + \sin{\phi} \left[2\frac{\Delta v}{v_{\rm k}} + \sin{\phi} \left(3 + \left(\frac{\Delta v}{v_{\rm k}} \right)^2 + 2 \sin{\phi} \frac{\Delta v}{v_{\rm k}}  \right)   \right]
.
\label{chieq}
\end{equation}

The possible extrema of equation (\ref{FirstPiece}) are, for inward panspermia,

\begin{equation}
\left(\phi\right)_{\rm ex} = \Bigg\lbrace \pm \frac{\pi}{2}, \ \ \
\pm \arccos{\left[\pm \frac{\sqrt{a^2 - q'^2}}{a}\right]}, \ \ \ 
  \pm \arccos{\left[\pm \frac{\sqrt{2a^3 + 2a^2 q' + 2a q'^2 + q'^3} }
      {a\sqrt{2a+q'}}\right]} \Bigg\rbrace
  .
\label{curveex}
\end{equation}

\noindent{}For outward panspermia, a similar set of extrema hold (just substitute $q'$ for $Q'$) except for the middle ones, where the radicand would always be negative. The resulting minimum, real and nonzero values of $\Delta v/v_{\rm k}$ can then be converted into a minimum kick speed. For outward panspermia, the only real solutions are $(\phi)_{\rm ex} = -\pi/2$.

\subsection{Probability distributions}

In order to obtain the probability distribution functions for the debris to intersect the orbit of a planet, we first invert 
equations (\ref{FirstPiece})-(\ref{SecondPiece}) as an explicit function of $\sin{\phi}$ and write this expression in terms 
of known dimensionless ratios. Doing so yields

\begin{equation}
\left(\sin{\phi}\right)_{\rm in} = 
\frac{
\alpha^2 - 1 \pm \sqrt{\alpha \left[ 2 + \alpha^3 + \alpha \left(\gamma^2 - 3\right) \right]  } 
}
{\gamma},
\label{sinin}
\end{equation}

\begin{equation}
\left(\sin{\phi}\right)_{\rm out} = 
\frac{
1 - \beta^2 \pm \sqrt{2\beta^3 + 1 + \beta^2 \left(\gamma^2 - 3\right)  } 
}
{\gamma \beta^2}
\label{sinout}
\end{equation}

\noindent{}where

\begin{equation}
\alpha \equiv \frac{q'}{a} < 1
\end{equation}

\begin{equation}
\beta \equiv \frac{a}{Q'} < 1,
\end{equation}

\begin{equation}
\gamma \equiv \frac{\Delta v}{v_{\rm k}}
.
\label{gammaeq}
\end{equation}

\noindent{}Equations (\ref{sinin}) and (\ref{sinout}) produce physical results 
only if the absolute value of the right-hand-sides are less than or equal to unity. 

\paragraph{Outward panspermia}

For outward panspermia -- the simpler case -- this condition
is consistent with those found in equations (\ref{useful1}-\ref{useful2}) for 
only the {\it lower} sign in equation (\ref{sinout}). That equation therefore
gives two values of $\left(\sin{\phi}\right)_{\rm out}$, denoted as $\phi_{1}^{\rm out}, \phi_{2}^{\rm out}$, 
as long as $\gamma$ is not sufficiently large or small. Assume over the interval
$[0,2\pi)$ that $\phi_{1}^{\rm out} \le \phi_{2}^{\rm out}.$ Consequently, for outward
panspermia, we obtain

\begin{itemize}

\item {\bf In the case of low kick speeds: if}

\begin{equation}
\gamma < \sqrt{\frac{2}{1+\beta}} - 1, \ \ \ \ {\rm then} \ P_{\rm out} = 0.
\label{out1}
\end{equation}

\item {\bf In the case of high kick speeds: if}

\begin{equation}
\gamma \ge \sqrt{\frac{2}{1+\beta}} + 1, \ \ \ \ {\rm then} \ P_{\rm out} = 1.
\end{equation}

\item {\bf In the case of medium kick speeds: if}

\begin{equation}
\sqrt{\frac{2}{1+\beta}} - 1 \le \gamma < \sqrt{\frac{2}{1+\beta}} + 1, \ \ \ {\rm then}
\nonumber
\end{equation}

\begin{equation}
P_{\rm out} = 
\frac{1}{2\pi} \left\lbrace 2\pi \left\lfloor \frac{\phi_{1}^{\rm out}}{\pi} \right\rfloor 
+ \left(-1 \right)^{\left\lfloor \frac{\phi_{1}^{\rm out}}{\pi} \right\rfloor}   \left(\phi_{2}^{\rm out} - \phi_{1}^{\rm out}\right)  \right\rbrace.
\label{out3}
\end{equation}

\end{itemize}

\noindent{}This last equation represents visually the normalized horizontal distance between points on a curve from Fig. \ref{aposample}. The potential modulation by $2\pi$ depends on whether or not this distance lies leftward or rightward of $\phi = \pi$.

\paragraph{Inward panspermia}

Now we perform a similar analysis for inward panspermia, which is more subtle. 
Consider again equation (\ref{sinin}). The condition that the absolute value of the
right-hand-side of the equation is less than or equal to unity now does not exclude either
sign in front of the radicand. 

The equation with the upper sign is valid when $\gamma$~$\ge$~$1 - \sqrt{2\alpha/(1+\alpha)}$
and the equation with the lower sign is valid when $\gamma \ge 1 + \sqrt{2\alpha/(1+\alpha)}$.
The latter accounts for the ``bump'' seen around $\phi = 270^{\circ}$ on Fig. \ref{perisample}.
The former specifies points along the sloped portions of the other curves. The vertical lines
on the figure correspond to extrema of $\phi$. 

Now let $\phi_{1}^{\rm in}$ and $\phi_{2}^{\rm in}$ be the solutions of equation (\ref{sinin}) with the upper sign
which respectively are closest to the extremum angles given by $\left[-\arccos{(-\sqrt{1-\alpha^2})} \right]$
and $\left[-\arccos{(\sqrt{1-\alpha^2})} \right]$. Further, let $\phi_{3}^{\rm in}$ and $\phi_{4}^{\rm in}$ be the
solutions of equation (\ref{sinout}) with the lower sign, such that $\phi_{3}^{\rm in} \le \phi_{4}^{\rm in}$ on
a $[0,2\pi)$ interval. Consequently, we obtain

\begin{itemize}

\item {\bf In the case of low kick speeds: if}

\begin{equation}
\left[\gamma < 1 + \sqrt{\frac{2\alpha}{1+\alpha}} \ \ {\rm and} \ \ \alpha < \frac{1}{2}\left(\sqrt{2} - 1\right)\right] \ \ {\bf or}
\nonumber
\end{equation}

\begin{equation}
\left[\gamma < \frac{1-\alpha}{2\alpha\left(1 + \alpha\right)} \ \ {\rm and} \ \ \alpha \ge \frac{1}{2}\left(\sqrt{2} - 1\right)\right],
\nonumber
\end{equation}

\begin{equation}
{\rm then} \ \ \ P_{\rm in} = 0.
\label{In1}
\end{equation}

\item {\bf In the case of medium kick speeds and wide separations, if}

\begin{equation}
1 + \sqrt{\frac{2\alpha}{1+\alpha}} \le \gamma < \frac{1-\alpha}{2\alpha\left(1 + \alpha\right)}
\ \ {\rm and} \ \ \alpha < \frac{1}{2}\left(\sqrt{2} - 1\right),
\nonumber
\end{equation}

\begin{equation}
{\rm then} \ \ \ P_{\rm in} = \frac{\phi_{4}^{\rm in} - \phi_{3}^{\rm in}}{2\pi}.
\end{equation}

\item {\bf In the case of medium kick speeds and close separations, if}

\begin{equation}
\frac{1-\alpha}{2\alpha\left(1 + \alpha\right)} \le \gamma < 1 + \sqrt{\frac{2\alpha}{1+\alpha}}
\ \ {\rm and} \ \ \ \alpha \ge \frac{1}{2}\left(\sqrt{2} - 1\right),
\nonumber
\end{equation}

\begin{equation}
{\rm then} \ \ \ 
P_{\rm in} = 
\frac{1}{2\pi} 
\left[
2 \left(-\arccos{\left(-\sqrt{1 - \alpha^2}\right)}   - \phi_{1}^{\rm in} \right)
\right].
\label{In3}
\end{equation}

\item {\bf In the case of high kick speeds: if}

\begin{equation}
\left[\gamma \ge \frac{1-\alpha}{2\alpha\left(1 + \alpha\right)}
\ \ {\rm and} \ \ \alpha < \frac{1}{2}\left(\sqrt{2} - 1\right)\right] \ \ {\bf or}
\nonumber
\end{equation}

\begin{equation}
\left[\gamma \ge 1 + \sqrt{\frac{2\alpha}{1+\alpha}}
\ \ {\rm and} \ \ \alpha \ge \frac{1}{2}\left(\sqrt{2} - 1\right)\right], \ \ {\rm then}
\nonumber
\end{equation}

\begin{equation}
P_{\rm in} = 
\frac{1}{2\pi} 
\left[
2\left(-\arccos{\left(-\sqrt{1 - \alpha^2}\right)}  - \phi_{1}^{\rm in}\right) 
+\phi_{4}^{\rm in} - \phi_{3}^{\rm in}
\right].
\label{In4}
\end{equation}

\end{itemize}

\noindent{}Equations (\ref{In1})-(\ref{In4}) all represent normalized horizontal distances between a curve on Fig. \ref{perisample}. Because the curves feature multiple extrema, equations (\ref{In3})-(\ref{In4}) in particular may not be obvious, but are derived from the same principle: if and where there are intersections between a curve and a horizontal line on the plot.

\section{Appendix B: A fiducial packed system}

The motivation for the formalism presented in this paper is compact planetary systems that are
within the habitable zones of their parent stars.  Compact systems are said to be ``packed''. 
The tendency for some systems to be packed
was hinted at a decade ago \citep{barray2004,baretal2008,rayetal2009}, but has now become 
commonplace thanks to the {\it Kepler} mission. In this appendix, we generate a compact planetary
system which allows one to quickly compute characteristic values without relying on system-specific data.

In order for compact systems to be dynamically stable, the planets must be sufficiently
separated from one another. Because no exact analytical prescription exists for three or more planets, many investigations have adopted empirical expressions. A commonly-used formulation
is to express the separation of two planets through equations (B1-B2) of \cite{chaetal2008} as

\begin{equation}
a_{u} = a_{u-1} + K \left[\frac{M_{u-1} + M_{u}}{3M_{\rm star}}\right]^{1/3}
         \left( \frac{a_{u-1} + a_{u}}{2} \right)
.
\label{recurs}
\end{equation}

\noindent{}Here $K$ is a constant and the planet subscripts are ordered in increasing distance
from their star. The larger the value of $K$, the longer the system will remain stable. However, as
$K$ increases, the scatter in the instability times becomes larger as well. For three-planet systems
$K \ge 5.5$ generally ensures that planetary systems will remain stable for at least $10^9$ yr.

If all of the planet masses are equal, then the recursive 
equation (\ref{recurs}) can be solved for arbitrary $u$, yielding

\begin{equation}
a_{u} = a_1 \left[ \frac{2 + K\mathcal{M}}{2 - K\mathcal{M}} \right]^{u-1}, \ \ u \ge 1
\end{equation}

\noindent{}where 

\begin{equation}
\mathcal{M} \equiv \left[\frac{2M}{3M_{\rm star}}\right]^{1/3}
\end{equation}

\noindent{}Therefore, in an $N$-planet packed system with equal planet masses and circular orbits,
the ratios $(a/Q')$ and $(a/q')$ are given just by a function of the constants $K$, $\mathcal{M}$
and the difference of the order numbers of the source and target planets. For example, suppose
the source is planet $j$ and the target, which is further away, is planet $k$. Then

\begin{equation}
{\rm min} \left(\Delta v \right)_{jk}^{\rm out} = \sqrt{\frac{G\left(M_{\rm star} + M\right)}{a_1}}
           \left[ \frac{2 + K\mathcal{M}}{2 - K\mathcal{M}} \right]^{\frac{1-j}{2}}
\sqrt{2 \left(1 + \left[ \frac{2 + K\mathcal{M}}{2 - K\mathcal{M}} \right]^{j-k} \right)^{-1} - 1} 
\label{outRecu}
\end{equation}

Consider, for example, 
a set of $N=6$ Earth-mass planets orbiting a Solar-mass star. Then, $\mathcal{M} = 0.0126$. Also assume
that $K = 7$ provides sufficiently distant spacing for the planets to remain stable
long enough to develop intelligent life. Consequently, $(2+K\mathcal{M})/(2-K\mathcal{M}) = 1.092$. The escape speed from each planet is 11.19 km/s, and the circular speed of the innermost
planet is [94.19 km/s $\times (a_1/0.10 {\ {\rm au}})^{-1/2}$]. Note that with this last relation,
all of the minimum thrusts computed may be scaled accordingly to the innermost semimajor axis.
For $a_1 = 0.10$ au, the minimum kick speeds needed to reach the outermost planet from the other planets starting with the innermost are then $\left\lbrace 43.9, 37.7, 31.3, 24.5, 16.6 \right\rbrace$ km/s.

\end{document}